# The Splice Bootstrap

Gerard Keogh*

*Trinity College Dublin.*

Abstract: This paper proposes a new bootstrap method to compute predictive intervals for nonlinear autoregressive time series model forecast. This method we call the splice boobstrap as it involves splicing the last p values of a given series to a suitably simulated series. This ensures that each simulated series will have the same set of p time series values in common, a necessary requirement for computing conditional predictive intervals. Using simulation studies we show the methods gives 90% intervals intervals that are similar to those expected from theory for simple linear and SETAR model driven by normal and non-normal noise. Furthermore, we apply the method to some economic data and demonstrate the intervals compare favourably with cross-validation based intervals.

## 1. Introduction

In time series analysis, a out of sample forecast of an unknown future value is called a point prediction and its distribution is known as the predictive distribution. When data are not normal this distribution plays an important role. The statistical uncertainty of the point prediction based on the predictive distribution is referred to as the future prediction interval, or simply the prediction interval. This interval will contain the future value with high probability; say 95% or 99%. In this study realistic prediction intervals for some threshold time series models driven by non-normal errors are computed using a novel adaptation of existing sieve type parametric bootstrap methods. This method is called the Time Series Splice Bootstrap or Splice Bootstrap (SB) for short.

For a fairly simple stationary nonlinear time series model with normal errors the so-called Normal Forecasting Error Method (see de Bruin 2002) can be used to analytically compute the predictive interval. Typically, with normal errors the predictive interval is given as the mean plus or minus a multiple of the standard deviation of the predictive distribution (e.g. 2 standard errors). When a time series model is nonlinear and the noise asymmetric and/or multimodal then analytic predictive intervals can be constructed by solving the forward Chapman-Kolmogorov intrgral equations (see Tong 1990, Ch's 4 and 6). Needless to say these equations quickly become intractable and one then must resort to computational methods. Tong (1990) adopts the obvious computational solution of Gauss-quadrature to compute the complex nonlinear integrals involved. While this to works for small time models the curse of dimensionality soon renders the approach impractical. To avoid these difficulties we propose in this paper an alternative approach based on bootstrap re-sampling (see Efron & Tibshriani 1990).

*email: gmkeogh@justice.ie

Bootstrap methods have been used with success for computing intervals for parameter estimates (see Buhlmann 2002) and autocorrelations (see Romano & Thombs, 1996). Intervals for the linear autoregressive model of order p, the AR(p) model driven by non-Normal disturbances are constructed in Thombs & Schucany (1990), Breidt et. al. (1995), Romano & Thombs (1996), Hansen (1999) and Kim (2002). The parametric bootstrap set out in Thombs & Schucany (1990) is the starting point of this research. In particular they use the fact that the AR(p) model driven by normal errors is reversible (see Box & Jenkins 1976) to create a replicate backcast series using just last p values from the given time series data as starting values. This procedure can be repeated many times to generate many replicate series. Each one of these is used in turn to generate a set of parameter estimates and these estimates along with the last p series values are used to generate a sequence of future prediction values. The distribution of these predicted values at each future time point is the empirical predictive distribution at that future time point.

The back casting device adopted Thombs & Schucany (1990) is both appealing and simple. Nonetheless it is only appropriate for linear models with symmetric error distributions such as the normal. In the presence of asymmetry the forward and backward predictive distributions are different and accordingly backcasting cannot be utilised. In this paper we present a novel extension to the parametric bootstrap where the back casting device of is avoided. In particular this renders the method suitable for asymmetric nonlinear time series models. The method is parametric and is called the Splice Bootstrap (SB).

The basic idea behind the SB is to fit say a p-th order (nonlinear) model to the given time series data and compute parameter estimates and innovation errors. Select an arbitary starting time point (or p points) from the given series and begin to generate a new value using the estimates and a value randomly selected from the innovation errors. Continue in this way until a very long time series is generated. Then sequentially search through the generated series to find the subsequence of values closest to the last p data series values. The point (or set of p) closest matching point is the splice point. At this point take a subseries of previous values of sufficient length from the generated and simply append or splice the last p values from the data series to the subseries. Re-estimate the model parameters from this series and use these to produce a sequence of future values. The basic idea is that the distribution of the spliced series will be close to that of the original data series and will have the same last p time series value if the discontinuity at the splice point is small. Importantly, only the forward innovation error distribution is re-sampled so there problem of backcasting is avoided.

This paper is organised as follows. In section 2 we review the theory and terminology around time series prediction intervals. In section 3 we set out the parametric bootstrap of Thombs & Schucany (1990) and use it as a basis to describe our SB. Section 4 examines how the SB performs on data simulated from simple time series models driven by different forms of errors. These studies show that the predictive intervals are accurate and consistent with increasing sample size for models driven by noise from normal, exponential and normal mixture models respectively. In section 5 assess how well the methods perform on a small number of economic time series that incorporate independent and seasonal effects. We contrast SB performance against cross-validation forecat error estimates over a forecast horizon of 12 steps (equivalently 1 year for monthly data). In section 6 we draw some conclusions.

## 2. Prediction Intervals

The overall objective is to obtain a prediction interval for a future value for a nonlinear auroregressive time series model of lag order p, an NLAR(p) model (see Tong 1990). Specifically we assume additive errors so the model for the time series value $y_t$ has the following general form

$$y_t = f(y_{t-1},\dots,y_{t-p}) + \varepsilon_t \tag{1}$$

where $f(\bullet)$ is a general nonlinear functional form over p lagged values of $y$ and $\varepsilon_t$ is a random shock at time $t$. This model can be used to be used to give a one-step-ahead autoregressive forecast

$$\hat{y}(T,1) = E(y_{T+1}|I_T) = f(y_T,\dots,y_{T-p+1}) \tag{2}$$

where $E(y_{T+1}|I_T)$ is the expectation of $y_{T+1}$ conditional on the information set $I_T$, comprising the lagged values $y_T, y_{T-1}, y_{T-2},\dots, y_{T-p}$ and the specified $f(\bullet)$. When this forcast value is used to compute the next forecast value $\hat{y}(T,2)$ at a future time $T+2$ according to

$$\hat{y}(T,2) = E(y_{T+2}|I_T) = f(\hat{y}(T,1), y_T,\dots, y_{T-p+2})$$

we get a so-called to plug-in forecast. Repeating this process $k$ times we can generate a sequence of $k$ plug-in forecasts. This "plug-in" method is the most common method adopted in practice.

Unfortunately, while plug-in forecast are very easy to compute they are inefficient. For example, a cross validation approach based on holding back the last $k$ values from a simulated time series can be adopted to compute the predictive distribution (see Keogh 2006, Ch 5). Given a set of $s$ simulated time series this procedure computes the predictive distribution $p(y_{n+k} \mid p(y_n))$ over this set of simulated time series. However, this distribution is not the true conditional predictive distribution, since $y_n$ is not fixed for every simulated dataset. As a consequence the cross validation approach computes an approximation to the unconditional or marginal distribution as $s \to \infty$ and clearly this may diverge from the predictive distribution as $k$ increases.

The (true) conditional predictive distribution at $k$ steps ahead given an infinite realisation $\mathbf{Y}_n = y_n, \ldots$ is

$$p(y_{n+k} \mid \mathbf{Y}_n) = \int_{-\infty}^{\infty} p(y_{n+k} \mid y_{n+1})\, p(y_{n+1} \mid \mathbf{Y}_n)\; dy_{n+1} \quad k = 1 \ldots \tag{3}$$

This is a recursive equation for the conditional predictive density based on solving the forward Chapman-Kolmogorov (C-K). As noted in the introduction Tong (1990 subsection 4.2.4.3) solves this equation using Gaussian quadrature. The distribution the resulting predictive interval, denoted by $PI_{n+k}$, that covers a future value $y_{n+k}$ with probability $\beta = 100(1-\alpha)\%$ is

$$PI_{n+k}(L_n, U_n) = P(L_n < y_{n+k} \le U_n \mid \mathbf{Y}_n)$$

where $L_n$ and $U_n$ are the upper and lower quantiles of the interval and $P(L_n < y_{n+k} \le U_n \mid \mathbf{Y}_n)$ is the corresponding distribution function. If, in addition, as S $\to \infty$ $E\{PI_{n+k}(L_n, U_n)\} \to 1-\alpha$ then $PI_{n+k}(L_n, U_n)$ is also a $\beta = 100(1-\alpha)\%$ unconditional interval for $y_{n+k}$. Predictive intervals constructed in this was are known as parametric intervals – the standard normal interval (e.g. Box-Jenkins 1976) is the most commonly used. Clearly, for autoregressions, the distribution of $y_{n+k} \mid \mathbf{Y}_p$ where $\mathbf{Y}_p = y_n, y_{n-1} \ldots y_{n-p+1}$ is the same as the distribution of $y_{n+k} \mid \mathbf{Y}_n$. Therefore it makes sense to write the conditional predictive interval for autoregressive processes as

$$PI_{p,n+k}(L_n, U_n) = P(L_n < y_{n+k} \le U_n \mid \mathbf{Y}_p) \tag{4}$$

The key purpose of this paper is to construct intervals of the type (4) and to show by simulation that $E\{PI_{p,n+k}(L_n, U_n)\} \to 1-\alpha = \beta$. In simulation studies we take $\alpha = 0.1$ and so we consider 90% intervals – we mention that simulation results for other values of $\beta$ not reported here are similar to the $\beta = 90\%$ case.

In this paper intervals are computed using the SB which is a type of parametric bootstrap approach. More generally the parametric bootstrap relies on finding $L_n$ and $U_n$ based on a parametric model for the data and then 'pivoting' on these critical values to isolate $y_{n+k}$. The SB method implemented here follows along similar line to Thombs & Schucany (1990) in that future values are used to define the root quantity $R_n = y_{n+1}$ (see Breidt et. al 1995). Letting $\hat{f}(y_n)$ denote the nonlinear model fit to the data, the root may be written as

$$R_n = \hat{f}(y_n) + \{y_{n+1} - \hat{f}(y_n)\} = \hat{f}(y_n) + \hat{\varepsilon}_t$$

where $\hat{\varepsilon}_t$ are the residuals from the model fit. This root quantity can then be used to generate replicate samples of the observed time series as

$$R_t^* = \hat{f}(y_t^*) + \hat{\varepsilon}_t^*$$

where $y_t^*$ and $\hat{\varepsilon}_t^*$ are replicates of $y_t$ and $\varepsilon_t$ respectively. The predictive interval (4) is then computed from this replicate series. In practice a large number (say 100) bootstrap replicate series are used to estimate the predictive interval.

The Thombs & Schucany (1990) method of constructing the predictive interval is described in detail in the next section as the SB is a nonlinear generalisation of their method . In general parametric bootstrap methods preserve the underlying characteristics of the data generating process by preserving correlation structure in the observations.

## 3. The Splice Bootstrap Method for Time Series

This section describes the Splice Bootstrap (SB). The section begins with a review of the parametric bootstrap method for computing predictive intervals of the linear AR(p) model given in Thombs & Schucany (1990). We discuss the limitations of their method in the context of nonlinear models. With a view to addressing these limitations the SB is then set out and its advantages for computing NLAR(p) intervals discussed.

### Linear AR(p) Parametric Bootstrapping

The Thombs & Schucany (1990) method is set out here for the AR(1) model; generalisations to AR(p) are straightforward. Consider the stationary AR(1) time series model defined by

$$y_t = \phi_f \, y_{t-1} + \varepsilon_t \tag{5a}$$

where $\phi_f$ is an unknown constant, $\{\varepsilon_t\}$ is a sequence of zero mean independent errors with common distribution function $F_\varepsilon$ having finite 2nd order moments and $t = 0, \pm1, \pm2, \ldots$. Model (5a) is called the forward model and associated with it is the backward model where

$$y_t = \phi_b \, y_{t+1} + e_t \tag{5b}$$

These two models have the same correlation structure (see Box & Jenkins 1976) endowing the time series with a useful time-reversible property. This property is particularly useful as it allows replicate series to be generated that have the same last value (or last p valuesfor an AR(p) model) and, in addition, have the same correlation structure. Clearly, from the definition of the predictive interval (4) replicate series having the same last value is a fundamental requirement.

The bootstrapped prediction interval of Thombs & Schucany (1990) involves computing both forward model centred residuals $\{\hat{\varepsilon}_t\}$ with distribution function $\hat{F}_\varepsilon$ from (5a) and backward centred residuals $\{\hat{e}_t\}$ with distribution function $\hat{F}_e$ from (5b). Starting with the last data value $y_n^* = y_n$ the estimated backward model is used to compute

$$y_{t-j}^* = \phi_b \, y_{t-j+1}^* + \hat{e}_{t-j}^*$$

where $\hat{e}_{t-j}^*$ are random i.i.d. draws from $\hat{F}_e$. Using the resulting bootstrap replicate series a new forward model is estimated giving $\hat{\phi}_f$. Future values $y_{n+k}^*$ are now computed using the new forward model parameter estimates and forward residuals drawn i.i.d. from $\hat{F}_\varepsilon$; note, these will be conditional on $y_n^* = y_n$. If we denote be the cdf of the future value $y_{t+k}^*$ by $G_B^*$ then the endpoints of the prediction interval are given by the quantiles of $G_B^*$. A prediction interval constructed in this way is a percentile interval in the sense described in Hall (1992).

In general there are two concerns with the above method. The first lies in the fact that the forward $F_\varepsilon$ and backward $F_e$ residual distributions are not the same when the innovation is correlated. As a consequence the re-sampling scheme should only be used when the innovation sequence is i.i.d. (see Thombs & Schucany 1990). Fortunately for NLAR(p) models of the form (1) this is the case as the innovation is assumed i.i.d. A second concern is that the method relies on the assumption of time reversibility. This cannot be assumed for NLAR(p) models. For example, even a simple asymmetric model such as the Sefl-Exciting Threshold Autoregressive (SETAR) of Tong (1990) has a correlation structure that is regime dependent and accordingly back casting does not apply.

**Nonlinear AR(p) Parametric Bootstrapping**

In this subsection a novel alternative to back casting is proposed that guarantees the last p values across every replicate series are the same. This involves removing the last $p$ values from the observed time series and appending them to end of each replicate series (of length $n-p$). If this is done in a sensible manner, then the distribution of future values will be conditional on these last $p$ values of the data.

Splice Bootstrap

1. Given an observed time series $y_t$ $(t=1,2,\dots,n)$ estimate the NLAR(p) model $f_t(\bullet)$ and compute the estimated innovation errors $\hat{e}_t = y_t - f_t(\bullet)$. The prediction interval will be computed for this model. This model also fixes the lag order $p$. The last $p$ values of the observed series are retained for appending to each replicate series.
2. Set m = 1,000.
3. Start with $y^*_{-m}, y^*_{-m+1} \dots, y^*_{-m+p-1}$ as a randomly selected subseries from $y_t$ and simulate $y^*_t$ for $t=-m+p,\dots,0,1,2,\dots,l, l>n$ from $f_t(\bullet)$ as $y^*_t = f_t(\bullet)+\hat{e}_t$ where $\hat{e}_t$ is a random draw from $\hat{F}_e$ the distribution of the estimated innovation errors rescaled by $\sqrt{(n-p)/(n-2p)}$ (see Thombs & Schucany 1990) to compensate for deflation in the innovation variance due model fitting.
4. Select the splice point $r > n-p$ such that $\left|y_{n-p+1} - y^*_r\right|, \left|y_{n-p+2} - y^*_{r+1}\right|, \dots \left|y_n - y^*_{r+p-1}\right|$ is a minimum.
5. Select the subseries of length $n-p$ from $\{y^*_t\}$ as $y^*_{r-n}, y^*_{r-n+1} \dots, y^*_r$.
6. To then end of this subseries splice the last $p$ values of the original series giving the splice bootstrap (SB) replicate series $y^*_{s-n}, y^*_{s-n+1} \dots, y^*_s, y_{n-p+1,\dots,} y_n$ of length $p$.
7. Estimate the NLAR(p) SB model $f_t^{SB}(\bullet)$ and compute the innovation errors $\hat{e}^*_t = y^*_t - f_t^{SB}(\bullet)$. Here the sieve bootstrap replicate series is used to estimate the TSMARS model $f_t^{SB}(\bullet)$.
8. TSMARS models that do not have the same form as the original model are rejected as invalid.
9. Future values are computed using the plug-in rule (see Chapter 5), the last $p$ original values and a valid TSMARS sieve bootstrap model $f_t^{SB}(\bullet)$. The forecast is given by $y^*_{n+k} = f^{SB}_{n+k}(\bullet) + \hat{e}^*_{t+k}$ where $\hat{e}^*_{t+k}$ (k > 0) is a random draw from $\hat{F}_e$, the distribution of the estimated sieve bootstrap replicate innovation errors.
10. Repeat steps 3-10 until B bootstrap future values of each $y^*_{t+k}$ are available.
11. Let $G^*_B$ be the cdf of the future value $y^*_{t+k}$ then the endpoints of the prediction interval are given by the quantiles of $G^*_B$.

A couple of points are worth noting about this algorithm. First, the predictive set of last $p$ values from the original series is retained to generate the conditional predictive distribution. This set is used both in the bootstrap replicate series and to start off the (Markovian) forecast sequence in step 9. The resulting interval therefore approximates (4) and gives the unconditional interval as $B \to \infty$.

Second, two different sets of innovations are used. The original sequence $\hat{e}_t$ is used to generate the bootstrap replicate series while the bootstrapped sequence $\hat{e}_t^*$ is used in forecasting. This provides sufficient mixing for every bootstrap replicate series. Both of these points are designed to ensure that bootstrap replicate series and forecasts provide a good approximation to the underlying predictive stationary distribution. Moreover, it is reasonable to conjecture based on Theorem 3.1 of Thombs & Schucany (1990) that $y_{t+k}^* \rightarrow y_{t+k}$ in distribution provided the $f_t^{SB}(\bullet) \rightarrow f_t(\bullet)$ in probability. Simulation studies to be conducted in the next section indicate this conjecture to be true.

## 4. SB Predictive Intervals for Simulated Models

### Models and Testing Procedure

To ascertain the quality of the bootstrap methods outlined in the previous section, simulations studies are conducted based on the linear AR(1) model and the SETAR(2,1,1) model. Both models are considered under three different i.i.d. noise distributions. These distributions are taken from Thombs & Schucany (1990) and are normal, exponential and a mixture of normals respectively. The AR(1) model is

$$y_t = -0.8\, y_{t-1} + \varepsilon_t \quad \text{where } \varepsilon_t \begin{cases} N(0,1) \\ Exp(1) \\ N(-1,1)\, U < 0.9;\, N(9,1)\, U \geq 0.1 \end{cases} \tag{6}$$

where $U \sim \text{Uniform}[0,1]$. Simulations are conducted based on each of these three models with n = 25, 50 and 100 observations respectively.

The SETAR(2,1,1) model is

$$y_t = \begin{matrix} 0.7 y_{t-1} \\ 0.3 y_{t-1} \end{matrix} + \varepsilon_t \quad \begin{matrix} if\ y_{t-1} \leq 0 \\ if\ y_{t-1} > 0 \end{matrix} \tag{7}$$

with the same noise distributions as in (6) and n = 100, 250 and 500 respectively.

For each of these model combinations and sample sizes a time series realisation is generated. Forecast values are also generated for $k$ steps ahead. To estimate the probability content $\beta = 100(1-\alpha)\%$ and average length of interval for these series we adopt the following test procedure.

1. Simulate a series of length n according to a specific model combination and generate R=100 future values $y_{n+k}^R$ at each step ahead $k$.
2. Use the bootstrap procedure to obtain a 90% prediction interval $G_n^*$ based on $(L_n^*, U_n^*)$.
3. Estimate the conditional coverage by $(\beta_k^* = \#\{L_n^* \leq y_{n+k}^R \leq U_n^*\})$ and interval length $Len(k) = U_n^* - L_n^*$.

Steps 1 – 3 are repeated 100 times to get a collection of summary measures $\{\beta_k^*, Len(k)\}$ In the Table Appendix we report the average value of the conditional coverage and its standard error, and also the average interval length and it standard error for the models considered. We also mention that simulation studies based on fitting an exact model for the simulated series show that the SB produces efficient interval when the number of bootstrap replicate series is increased to 999.

**Discussion of Results**

For both models (6) and (7) the SB results for an 5-step ahead prediction horizon are displayed in Tables A.1 (see Table Appendix).

Focussing on the AR(1) model driven by exponential noise the prediction interval for the smallest sample size (n=25) appears a little disappointing as the conditional coverage falls short of 90% for both bootstrap methods. However, these figures compare well with the conditional coverage figures in Thombs & Schucany (1990) – these were lower than nominal but nevertheless within a few percentage points. Nonetheless, as the sample size is increased the conditional coverage approaches the nominal 90% showing the methods are asymptotically efficient. This is in keeping with expectations and moreover the accuracy of the intervals as n increases is equal to that obtained by Thombs & Schucany (1990).

The detailed table in the Appendix, Table A.1, shows that similar results are obtained for this AR(1) model with normal and normal mixture noise models respectively. It can be observed interval lengths at each step for normal noise are slightly narrower than their asymptotic counterparts of 3.3, 4.2 and 4.7 respectively. Taking all of these observations together we conclude that the SB gives accurate and consistent predictive intervals for this model.

Looking at the nonlinear SETAR(2,1,1) model driven the prediction intervals obtained for all sample sizes are credible. In particular, for normal errors the intervals compare favourably with the theoretical interval computed via the Chapman-Kolmogorov given in Keogh (2006, Ch 5.). Moreover, there is an improvement in the accuracy of the interval as the sample size increases providing evidence for asympototic efficiency.

In conclusion over a range of time series lengths for both linear and nonliner models driven by normal and non-normal errors the SB produces credible prediction intervals. Moreover, the coverage probabilities and interval lengths appear to converge to nominal values as the sample size increased. A particularly appealing feature of the SB is that the nonlinearity in the SETAR model is captured resulting in excellent coverage probabilities and interval lengths. Accordingly, we may postulate that the SB can be expected to produce reliable and consistent intervals for more general NLAR(p) model of the form (1).

### 5. Bootstrap Intervals for Short-term Economic Time Series

In this section the SB is applied to some of the test bed series. Bootstrap methods are only applicable when data are stationary. Therefore attention is focussed on models arising from the Seasonal TSMARS method, STSMARS (see Keogh 2006). The purpose of this section is to generate prediction intervals for those stationary STSMARS models found in Chapter 4 of Keogh (2006). These are contrasted against their cross-validation counterparts obtained in Chapter 5 of Keogh (2006). Any unexplained divergence between these will indicate that the bootstrap methodology is defective and render the prediction interval useless.

STSMARS takes a time series $y_t$ and after appropriate transformations gives the series $z_t$. The lagged predictors $z_{t-1}, z_{t-2}, z_{t-3}, z_{t-s}, z_{t-(s+1)}$ are then input into the TSMARS program (see Keogh 2006) along with appropriately differenced trading effects predictors. The maximum interaction degree is set to 3 and basis function threshold = 2 X $10^{-8}$. Bootstrap future values are generated for the transformed series $z_t$. On completion of the TSMARS call, the sequence of transformations are applied in reverse, giving predictions for the model (see Chapter 4 for definition of terms)

$$\hat{y}_t = \exp\left[(1-B)^{-d}(1-B^s)^{-D}\left\{f\left(z_{t-1}, z_{t-2}, z_{t-3}, z_{t-s}, z_{t-(s+1)}, z_{\mathrm{t,MD}}, z_{\mathrm{t,TD}}, z_{\mathrm{t,EASTER}}\right)\right\}\right] - c$$

Bootstrap prediction intervals are then computed from these predicted values.

There are two key differences between bootstrapping the simple models of the last section and STSMARS models. First, the independent trading day predictors have to be included. These are generated as fixed effects based on the date of the future values n+k (k=1,…12). They are reused in every bootstrap replicate to generate the set of future values. Specifically, they are included in the model $f_{n+k}^{SB}(\bullet)$ in step 9 of the SB. A second consideration is seasonality. This is not a problem for the Sieve Bootstrap as replicate series are generated directly from the STSMARS model.

Eight test bed series given in Table A.2 are used to generate predictive intervals for up to 12 steps ahead. These series are taken from the Irish Central Statistics Office databank. The statistics reported are the mean predictive value, its coefficient of variation and inter quartile range (divided by the median predicted value) denoted by Mean, CV and SIRQ respectively. The mean value actually quoted is the percentage difference of the mean predicted value from the mean value of the original data. This indicates how far the predictive value is away from the centre of the data. A 'good' starting value for forecasting purposes is a value that is close to the mean.

The results in Table A.2 show a good degree of consistency across all problems considered. In particular the predictive mean value is consistent. It does not tend to the mean value as the forecasting horizon increases, but remains stable reflecting the fact that the data are differenced.

There is also a good degree of consistency between the CV value at each step ahead and the SIRQ value. The latter figure, as expected, being generally larger.

The results in Table A.2 are useful, but of limited value. Of greater interest is their comparison with cross-validation forecast errors given in Keogh (2006, Ch 5 Table 5.5.1.1). The % average residual error given in Table 5.5.1.1 is compared with the average CV obtained (over the 12 –steps ahead) for each problem in Table A.2. These figures are displayed in Table 1.

Table 1: Comparison of Errors

| | Series Number | | | | | |
|---|---|---|---|---|---|---|
| | 1 | 2 | 11 | 13 | 19 | 20 |
| % Error in Table 5.5.1.1 | 13.2 | 82.5 | 13.0 | 10.2 | 37.9 | 11.4 |
| Sieve Bootstrap CV | 14.1 | 193.2 | 12.7 | 7.9 | 27.1 | 3.1 |

The figures in Table 1 show the SB tends to give % errors that are close to cross validation errors. For Problem 2, outliers near the end of the series induce nonstationarity. This widens the size of predictive interval greatly and shows the SB takes account of nonstationary behaviour near the end of this series, an appealing feature. We mention that this effect was also evident in Keogh (2006, Ch 5 Table 5.5.1.1) where the maximum error was 175% but cross-validation over smooths this effect producing narrow predictice intervals. Accordingly, the SB is shown to work well and give reasonable predictive distributions. This allied to the fact the method worked well for the simple simulated models, demonstrates that the bootstrapping predictive intervals can be relied on. It is also nice to see that the cross validation intervals are in line with the bootstrap figures in most instances.

## 6. Closing Remarks

Novel variations of the parametric sieve type have been set out and their merits tested on data simulated from simple time series models driven by normal innovations and on a number of empirical time series.

Three modifications to existing bootstrap scheme were implemented. First, both methods endeavoured to recreate the true predictive distribution by retaining the last (or last p) values of the series in every bootstrap replicate sample. Second, the methods ensured sufficient mixing is retained in the bootstrapped replicate series. This was accomplished using two distinct sequences of innovations; namely those from the original model to generate bootstrap samples, and those from the bootstrap model for forecasting. Third, only correct models were used to build forecasted values. With these three modifications, parametric bootstrap method for time series is adapted to compute predictive intervals for nonlinear models. Tests on simple simulated models found the methods produced efficient predictive intervals.

The methods were also applied to a subset of the test-bed problems. Once again reasonable predictive distributions were obtained. The predictive intervals were also compared to cross-validation based intervals. This comparison showed that the bootstrap intervals were similar. As a consequence the predictive intervals obtained are accurate and reliable estimates for these empirical series. Moreover, the efficiency demonstrated indicates the methods should work for other time series modelling methods.

The main limitation of this approach to generating the predictive distribution lie in the fact that splicing introduces a discontinuity While the effect of this discontinuity is minimised by finding a suitable point at which to splice the retained last p-values of the original series to the bootstrap replicate series, no effort has been made to assess the impact of the discontinuity on the predictive interval.

**Table Appendix**

Table A.1: 90% Bootstrapped forward prediction estimates for the AR(1) and SETAR(2,1,1) Model

| | AR(1) | | | | SETAR(2,1,1) | | | |
|---|---|---|---|---|---|---|---|---|
| Steps Ahead $k$ | Mean $\beta_k^*$ | % $S.E.(\beta_k^*)$ | Mean $Len(k)$ | % $S.E.(Len(k))$ | Mean $\beta_k^*$ | % $S.E.(\beta_k^*)$ | Mean $Len(k)$ | % $S.E.(Len(k))$ |
| Normal Noise | | | | | | | | |
| n=25 | | | | | n=100 | | | |
| 1 | 84.7 | 20 | 3.3 | 30 | 89.1 | 16 | 1.7 | 10 |
| 2 | 84.2 | 20 | 4.0 | 20 | 88.9 | 15 | 1.9 | 10 |
| 3 | 83.8 | 30 | 4.5 | 30 | 89.6 | 15 | 1.9 | 10 |
| 4 | 83.4 | 20 | 4.6 | 30 | 88.5 | 16 | 1.9 | 10 |
| 5 | 82.3 | 20 | 4.7 | 30 | 89.0 | 16 | 2.0 | 20 |
| n=50 | | | | | n=250 | | | |
| 1 | 88.1 | 10 | 3.3 | 10 | 89.1 | 10 | 1.7 | 10 |
| 2 | 87.4 | 10 | 4.1 | 10 | 89.1 | 5 | 1.9 | 10 |
| 3 | 86.6 | 10 | 4.6 | 20 | 88.6 | 10 | 1.9 | 10 |
| 4 | 85.9 | 10 | 4.8 | 20 | 87.6 | 10 | 1.9 | 10 |
| 5 | 86.6 | 10 | 5.0 | 20 | 88.7 | 10 | 2.0 | 10 |
| n=100 | | | | | n=500 | | | |
| 1 | 88.4 | 10 | 3.2 | 10 | 89.0 | 5 | 1.7 | 9 |
| 2 | 87.9 | 10 | 4.1 | 10 | 89.1 | 5 | 1.9 | 8 |
| 3 | 87.1 | 10 | 4.5 | 10 | 89.4 | 5 | 1.9 | 10 |
| 4 | 86.7 | 10 | 4.7 | 10 | 88.6 | 5 | 2.0 | 10 |
| 5 | 87.5 | 10 | 5.0 | 10 | 89.3 | 5 | 2.0 | 9 |
| Exponential Noise | | | | | | | | |
| Steps Ahead $k$ | Mean $\beta_k^*$ | % $S.E.(\beta_k^*)$ | Mean $Len(k)$ | % $S.E.(Len(k))$ | Mean $\beta_k^*$ | % $S.E.(\beta_k^*)$ | Mean $Len(k)$ | % $S.E.(Len(k))$ |
| n=25 | | | | | n-100 | | | |
| 1 | 86.4 | 20 | 3.0 | 40 | 82.5 | 10 | 1.5 | 37 |
| 2 | 82.4 | 30 | 3.7 | 40 | 83.6 | 10 | 1.5 | 40 |
| 3 | 80.7 | 30 | 4.1 | 40 | 84.0 | 10 | 1.5 | 40 |
| 4 | 80.5 | 30 | 4.3 | 40 | 84.0 | 10 | 1.6 | 41 |
| 5 | 78.7 | 30 | 4.3 | 40 | 84.6 | 10 | 1.6 | 42 |
| n=50 | | | | | n=250 | | | |
| 1 | 90.4 | 10 | 2.9 | 30 | 90.1 | 10 | 1.4 | 20 |
| 2 | 86.4 | 10 | 3.9 | 20 | 88.2 | 10 | 1.5 | 20 |
| 3 | 85.7 | 10 | 4.2 | 20 | 88.6 | 10 | 1.6 | 20 |
| 4 | 85.3 | 10 | 4.5 | 20 | 88.5 | 10 | 1.6 | 20 |
| 5 | 85.6 | 10 | 4.6 | 20 | 88.7 | 10 | 1.6 | 20 |
| n=100 | | | | | n=500 | | | |
| 1 | 90.3 | 10 | 2.9 | 20 | 89.3 | 10 | 1.5 | 20 |
| 2 | 87.2 | 10 | 4.0 | 20 | 89.4 | 30 | 1.6 | 10 |
| 3 | 86.7 | 10 | 4.3 | 20 | 89.3 | 20 | 1.6 | 10 |
| 4 | 86.4 | 10 | 4.6 | 20 | 88.8 | 10 | 1.5 | 10 |
| 5 | 86.2 | 10 | 4.7 | 20 | 88.6 | 10 | 1.5 | 10 |
| | Sieve Bootstrap | | | | Vectorised Block Bootstrap | | | |
| Steps Ahead $k$ | Mean $\beta_k^*$ | % $S.E.(\beta_k^*)$ | Mean $Len(k)$ | % $S.E.(Len(k))$ | Mean $\beta_k^*$ | % $S.E.(\beta_k^*)$ | Mean $Len(k)$ | % $S.E.(Len(k))$ |
| Normal Mixture Noise | | | | | | | | |
| n=25 | | | | | n=100 | | | |
| 1 | 87.2 | 20 | 10.5 | 40 | 73.4 | 49 | 4.8 | 54 |
| 2 | 84.7 | 20 | 14.3 | 40 | 73.3 | 48 | 5.1 | 52 |
| 3 | 83.0 | 20 | 14.7 | 40 | 72.8 | 48 | 5.2 | 53 |
| 4 | 81.5 | 20 | 15.1 | 40 | 72.7 | 47 | 5.3 | 52 |
| 5 | 81.6 | 20 | 15.4 | 40 | 72.3 | 47 | 5.4 | 52 |
| n=50 | | | | | n=250 | | | |
| 1 | 89.1 | 10 | 10.8 | 30 | 86.4 | 19 | 5.3 | 26 |
| 2 | 86.4 | 10 | 14.6 | 30 | 85.8 | 19 | 5.7 | 22 |
| 3 | 86.4 | 10 | 15.4 | 30 | 65.1 | 18 | 5.7 | 21 |
| 4 | 84.6 | 20 | 15.6 | 30 | 84.7 | 18 | 5.8 | 23 |
| 5 | 85.6 | 20 | 16.8 | 30 | 84.5 | 19 | 5.9 | 21 |
| n=100 | | | | | n=500 | | | |
| 1 | 89.6 | 7 | 10.8 | 25 | 89.0 | 7 | 5.6 | 17 |
| 2 | 86.6 | 8 | 15.0 | 27 | 86.8 | 9 | 5.7 | 14 |
| 3 | 86.9 | 7 | 15.8 | 20 | 87.5 | 7 | 6.0 | 12 |
| 4 | 87.0 | 7 | 16.4 | 18 | 86.8 | 8 | 5.9 | 15 |
| 5 | 87.3 | 7 | 16.8 | 19 | 88.1 | 6 | 6.1 | 14 |

## Table A.2: Bootstrapped forward prediction estimates for the selected Test Bed Series

| Statistic | 1 | 2 | 3 | 4 | 5 | 6 | 7 | 8 | 9 | 10 | 11 | 12 |
|---|---|---|---|---|---|---|---|---|---|---|---|---|
| | Series 1: Cows Milk Protein Content (%) $\Delta\left[y_t = -0.01 - 0.44\, y_{t-1} + 0.48\, y_{t-12}\right]$ | | | | | | | | | | | |
| Mean | 75.9 | 29.1 | 24.9 | 50.1 | 32.4 | 26.8 | 45.6 | 33.0 | 24.0 | 51.6 | 27.9 | 30.7 |
| CV | 0.0 | 12.0 | 12.4 | 15.6 | 16.4 | 15.3 | 15.3 | 12.0 | 13.0 | 13.3 | 15.5 | 14.5 |
| SIRQ | 0.0 | 13.1 | 17.5 | 15.3 | 20.1 | 23.2 | 19.8 | 15.8 | 15.6 | 18.5 | 19.5 | 21.5 |
| | Series 2: Calves Slaughtering 000 Heads $y_t = 1.08 + 0.18\, y_{t-1}(y_{t-2} - 1.2)_- - 1.2\, y_{t-1}(y_{t-2} - 1.2)_-(y_{t-12} - 2.1)_- + 0.5\, y_{t-1}(y_{t-2} - 1.2)_-(y_{t-12} - 0.8)_-$ | | | | | | | | | | | |
| Mean | -3.2 | -1.9 | -0.7 | -1.5 | -1.5 | -0.6 | -0.7 | -1.4 | -0.8 | -1.3 | -0.7 | -0.9 |
| CV | 0.0 | 100.0 | 129.4 | 153.7 | 129.1 | 136.3 | 153.4 | 180.3 | 255.5 | 319.5 | 261.1 | 307.0 |
| SIRQ | 0.0 | 111.1 | 177.4 | 282.9 | 123.6 | 141.1 | 207.4 | 195.6 | 255.1 | 236.3 | 261.3 | 272.7 |
| | Series 6: Live Register Total (No) $\Delta\Delta_{12}\left[y_t = -4442 + 0.39y_{t-1} + 1.25(y_{t-12} - 6772)_-\right]$ | | | | | | | | | | | |
| Mean | -40.5 | -33.5 | -34.0 | -35.0 | -40.3 | -35.5 | -32.8 | -26.6 | -25.4 | -33.1 | -36.9 | -37.5 |
| CV | 0.0 | 1.5 | 1.2 | 1.1 | 1.9 | 1.6 | 1.2 | 1.5 | 1.4 | 1.2 | 1.6 | 1.8 |
| SIRQ | 0.0 | 1.9 | 1.6 | 1.1 | 3.1 | 2.1 | 1.6 | 1.2 | 1.5 | 1.9 | 2.9 | 2.5 |
| | Series 9: Live Register/ Nenagh Males (No) $\Delta_{12}\left[y_t = -0.16 + 0.24y_{t-1} + 0.26y_{t-2} + 0.2(y_{t-13} - 0.14)_+ + 1.04y_{t-1}(y_{t-12} - 0.52)_+\right]$ | | | | | | | | | | | |
| Mean | -26.7 | -28.6 | -27.2 | -27.8 | -29.2 | -29.0 | -26.8 | -24.8 | -28.0 | -30.3 | -33.0 | -37.8 |
| CV | 0.0 | 3.7 | 4.2 | 5.2 | 4.1 | 4.6 | 5.5 | 3.0 | 3.6 | 4.0 | 6.3 | 7.5 |
| SIRQ | 0.0 | 6.5 | 6.3 | 5.9 | 5.7 | 8.7 | 5.6 | 5.5 | 7.2 | 3.5 | 7.0 | 11.0 |
| | Series 11: Volume Index NACE 37 (Base 1985= 100) $\Delta\left[y_t = 0.15 - 0.3y_{t-2} - 0.14y_{t-3} + 4.85(y_{t-1} - 0.06)_+ - 0.18(y_{t-1} - 0.06)_-\right]$ | | | | | | | | | | | |
| Mean | 53.0 | 28.4 | 44.7 | 38.0 | 41.3 | 51.5 | 35.5 | 51.0 | 63.3 | 61.5 | 66.1 | 70.5 |
| CV | 0.0 | 11.2 | 15.5 | 11.7 | 11.9 | 20.5 | 13.9 | 10.1 | 10.1 | 10.3 | 15.2 | 9.8 |
| SIRQ | 0.0 | 16.0 | 17.3 | 15.6 | 12.8 | 22.0 | 23.8 | 9.2 | 11.8 | 10.4 | 6.7 | 9.6 |
| | Series 13: Volume Index NACE 429 Adjusted (Base 1985= 100) $\Delta\left[y_t = 0.28 + 0.18y_{t-2} - 0.87(y_{t-1} - 1.28)_- + 0.2(y_{t-12} - 1.05)_+\right]$ | | | | | | | | | | | |
| Mean | -2.9 | -14.5 | -31.2 | -18.5 | -10.3 | -11.4 | -31.5 | -51.9 | -19.1 | 6.8 | -16.7 | -28.7 |
| CV | 0.0 | 7.6 | 7.3 | 7.8 | 7.8 | 7.9 | 9.7 | 6.1 | 7.1 | 8.3 | 8.8 | 8.5 |
| SIRQ | 0.0 | 7.7 | 10.1 | 8.8 | 9.7 | 10.4 | 10.7 | 10.1 | 7.7 | 10.6 | 11.3 | 9.9 |
| | Series 19: Imports SITC 71 Power Machinery €000 $\Delta_2\begin{bmatrix} y_t = -64.070 - 0.28y_{t-2} + 0.36y_{t-3} - 0.0001(y_{t-1} - 904{,}100)_+(y_{t-13} - 1{,}012{,}000)_- + \\ 0.0001(y_{t-1} - 904{,}100)_+(y_{t-2} - 204)_-(y_{t-13} - 1{,}012{,}000)_- \end{bmatrix}$ | | | | | | | | | | | |
| Mean | 17.9 | 50.7 | 17.3 | 16.1 | 12.0 | 32.8 | 5.8 | 20.9 | 46.3 | 26.0 | 22.4 | 29.0 |
| CV | 0.0 | 20.9 | 24.2 | 29.2 | 20.7 | 31.4 | 21.6 | 18.0 | 36.5 | 49.5 | 28.3 | 18.3 |
| SIRQ | 0.0 | 30.8 | 42.1 | 24.4 | 40.5 | 58.9 | 24.2 | 25.3 | 48.1 | 25.7 | 31.3 | 31.1 |
| | Series 20: Exports Adjusted €000 $\Delta_2\begin{bmatrix} y_t = -64.070 - 0.28y_{t-2} + 0.36y_{t-3} - 0.0001(y_{t-1} - 904{,}100)_+(y_{t-13} - 1{,}012{,}000)_- + \\ 0.0001(y_{t-1} - 904{,}100)_+(y_{t-2} - 204)_-(y_{t-13} - 1{,}012{,}000)_- \end{bmatrix}$ | | | | | | | | | | | |
| Mean | 63.4 | 61.2 | 55.6 | 61.2 | 57.2 | 51.5 | 54.8 | 47.8 | 54.7 | 51.6 | 50.4 | 52.2 |
| CV | 0.0 | 1.7 | 2.8 | 2.8 | 3.0 | 2.7 | 3.5 | 3.9 | 3.4 | 3.7 | 3.1 | 3.4 |
| SIRQ | 0.0 | 1.3 | 1.6 | 1.8 | 1.7 | 2.3 | 2.8 | 2.1 | 2.4 | 3.5 | 2.5 | 2.2 |